# Implementación de un sistema IoT de bajo costo para el monitoreo de la calidad del aire en El Salvador

Omar Otoniel Flores Cortez
Facultad de Informatica y Ciencias Aplicadas
Universidad Tecnologica de El Salvador
San Salvador, El Salvador
omar.flores@utec.edu.sv

Ronny Adalberto Cortez Reyes
Facultad de Informatica y Ciencias Aplicadas
Universidad Tecnologica de El Salvador
San Salvador, El Salvador
ronny.cortez@utec.edu.sv

Veronica Idalia Rosa
Facultad de Informatica y Ciencias Aplicadas
Universidad Tecnologica de El Salvador
San Salvador, El Salvador
veronica.rosa@utec.edu.sv

*Abstract*—La contaminación ambiental es un factor que representa un importante riesgo para la salud. En El Salvador, el ente encargado del monitoreo de la calidad del aire es el Ministerio de Medio Ambiente y Recursos Naturales, actualmente dicho ministerio solamente cuenta con 3 estaciones para el monitoreo de la calidad del aire en todo el territorio del país. El objetivo principal de este trabajo fue la aplicación de técnicas de internet de las cosas, sistemas embebidos y sensores electrónicos en el diseño e implementación de una estación de monitoreo remoto de contaminación por material de partículas en el aire circundante. Así como la configuración de una plataforma de Internet de las Cosas (IoT) para el despliegue de los datos colectados por la estación remota. El desarrollo metodológico de esta investigación se basó en el Modelo de Arquitectura de Referencia para sistemas IoT, que tiene como base el desarrollo de prototipos de sistemas basándose en la correcta elección de componentes disponibles y adecuados al entorno o aplicación específica. Para el caso se utilizaron como insumos para el hardware electrónico un controladorEsp32 junto con una tarjeta de desarrollo Wemos-Lolin junto a un sensor de contaminación por material de partículas PMS5003; por el lado del software para la plataforma o "nube" de IoT se implementó un sistema de almacenamiento, gráficos y sitio web basado en herramientas low-cost. El principal resultado obtenido en este trabajo fue un prototipo IoT de estación electrónica que permite monitorear los niveles de contaminación por material de partículas en el ambiente, cuyos datos son accesibles desde cualquier dispositivo con acceso a internet a través de un sitio web, otro de los resultados fruto de este trabajo es la configuración de una plataforma o nube de IoT para: la conexión inalámbrica con la estación electrónica, el almacenamiento de los datos producidos por esta y una etapa de visualización web. La plataforma web diseñada para visualizar los tableros y datos recolectados por las estaciones, pueden ser accedido desde https://tinyurl.com/utecAQ

## I. INTRODUCCION

El medio ambiente es esencial para el ser humano. Es importante conocer en tiempo real el comportamiento de algunas magnitudes como la temperatura, la humedad, la presión atmosférica, la velocidad del viento, la contaminación del aire porque algunos aspectos como la agricultura, el transporte, las comunicaciones, la salud y otros pueden verse afectados por cambios en cualquiera de las magnitudes antes mencionadas. Uno de los contaminantes ambientales de mayor impacto es el Material de Partículas, o PM por sus siglas en ingles. Este contaminante se define como una mezcla de partículas líquidas y sólidas, sustancias orgánicas e inorgánicas, que están suspendidas en el aire. Las partículas contaminantes se clasifican como PM10 (<= 10um) y PM2.5 (<= 2.5um).

Las denominadas PM10 son partículas inhalables que tienen diámetros de, por lo general, 10 micrómetros y menores. Las fuentes de emisión de estas partículas pueden ser móviles o estacionarias, destacando que un 77.9% de la cantidad total emitida de PM10 procede del polvo suspendido existente en la atmosfera. Como fuentes minoritarias de contaminación es importante señalar que el 3.7% del total procede de quemas agrícola y un 3.3% es de origen doméstico. [1]

En el caso del contaminante PM2.5 lo constituyen partículas inhalables finas que tienen diámetros de, por lo general, 2,5 micrómetros y menores. Las partículas finas pueden provenir de diversas fuentes. Incluyen centrales eléctricas, vehículos motorizados, aviones, quema de madera residencial, incendios forestales, quema agrícola, erupciones volcánicas y tormentas de polvo; algunos se emiten directamente al aire, mientras que otros se forman cuando los gases y las partículas interactúan entre sí en la atmosfera.

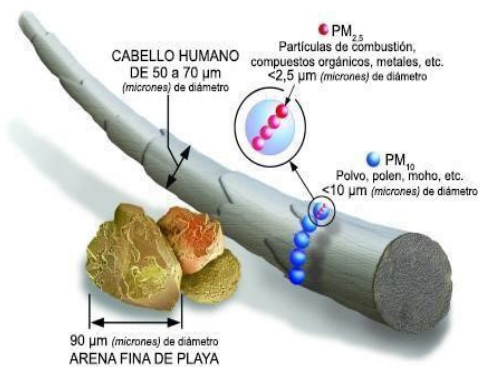

Fig. 1. Comparativa en dimensión entre el contaminante PM10 y PM2.5 versus un cabello humano y un grano de arena.

## A. Índice centroamericano de calidad del Aire ICCA

Es un índice numérico adimensional para la notificación de la calidad del aire a diario. Indica el grado de pureza o contaminación atmosférica y los efectos para la salud; gira en torno a los efectos en la salud que se pueden sufrir en unas cuantas horas o días tras respirar aire contaminado. El ICCA una escala numérica que va de 0 a 500, cuanto más alto es el valor del índice, mayor es el nivel de contaminación atmosférica y mayor la preocupación para la salud, se divide en sus niveles de preocupación por la salud, según la imagen a continuación. [1] Cada categoría corresponde a un nivel diferente de preocupación por la salud, además para mejor efecto visual ante la población se definen una escala de colores correspondiente, ver figura 2.

El ICCA se puede calcular tomando en consideración el valor del contaminante PM2.5. Cabe destacar que, en otras regiones como Norteamérica o Europa, el Índice de calidad del aire toma en consideración otros contaminantes además del PM2.5. Por lo tanto, en El Salvador para monitorear el ICCA este trabajo propuso el desarrollo de un sistema de IoT formado por estaciones electrónicas de medición del contaminante PM2.5 con capacidad de conexión Wifi junto con una plataforma en internet para la recolección y visualización delos datos producidos por estas.

Según la Organización Panamericana de la Salud (OPS), El Salvador es uno de los países que está catalogado sin capacidad para monitorear la calidad del aire. [2]. En El Salvador, el ente encargado del monitoreo de la calidad del aire es el Ministerio de Medio Ambiente y Recursos Naturales, MARN por sus siglas. Actualmente el MARM cuenta con tan solo 3 estaciones para el monitoreo de la calidad del aire en todo el territorio del país, denominada Red de Monitoreo de Calidad del Aire o REDCA, cabe destacar que estas estaciones solo de concentran en la ciudad capital San Salvador. [1]

Algunos de los problemas de las estaciones actuales son: alto costo de adquisición inicial, alto costo de mantenimiento o reparación, son voluminosas, a lo largo del tiempo han perdido características a tal grado que al día de hoy solo reportan medición de una variable contaminante, alta logística de instalación y puesta en marcha y poco o nulo personal capacitado experto.

Trabajos anteriores se han relacionado con el desarrollo de estaciones de monitoreo de la calidad del aire [3] [4]. Pero estos se han centrado en el uso de herramientas tecnológicas de alto costo. El hardware utilizado para la implementación es de tipo restringido o de código cerrado, y las plataformas en la nube para IoT son privadas y costosas. El uso de las llamadas nubes libres de IoT como *Thingspeak* o *Ubidots* es popular entre los estudios anteriores, pero tienen algunas limitaciones o están restringidos en número y frecuencia de telemetrías o número de datos que se pueden enviar en un determinado periodo de tiempo. [5] [6]

## B. Internet de Las Cosas

Internet de las cosas es un sistema de dispositivos de computación interrelacionados, con la capacidad de transferir datos a través de una red, sin requerir de interacciones humano a humano o humano a computadora.

| ICCA | Calidad del aire Material particulado (PM) | Contaminante (μg/m³) | | Indicaciones para su salud |
|---|---|---|---|---|
| | | PM2.5 | PM10 | |
| 0 - 50 | Buena | 0- 15.3 | 0- 54 | No se anticipan impactos a la salud cuando la calidad del aire se encuentra en este rango. |
| 51 - 100 | Moderada | 15.5-40.2 | 56-154 | No se anticipan impactos a la salud cuando la calidad del aire se encuentra en este rango. |
| 101 - 150 | Dañina a la Salud de los Grupos Sensibles | 40.5-65.4 | 155-254 | Los niños y adultos activos, y personas con enfermedades respiratorias tales como el asma, deben evitar los esfuerzos físicos excesivos y prolongados al aire libre. |
| 151 - 200 | Dañina a la Salud | 66-159 | 255-354 | La gente con la enfermedades respiratorias tal como asma, debe evitar el esfuerzo al aire libre; todos los demás, especialmente los mayores y los niños, deben limitar el esfuerzo prolongado al aire libre. |
| 201 - 300 | Muy dañina a la Salud | 160-250 | 355-424 | La gente con enfermedades respiratorias tal como asma, debe evitar todo el esfuerzo al aire libre; especialmente los mayores y los niños, deben limitar el esfuerzo prolongado al aire libre. |
| 301 - 500 | Peligroso | 251-500 | 424-604 | Todos deben evitar el esfuerzo al aire libre; gente con la enfermedad respiratoria tal como asma, debe permanecer dentro |

Fig. 2. Escala del ICCA con sus respectivas afecciones y colores

Un sistema de IoT involucra múltiples niveles o bloques, cada nivel debe estar diseñado para diferentes funciones y sus propios componentes y configuración. El diseño de los sistemas de IoT puede ser una tarea compleja y desafiante, ya que implica la interacción entre dispositivos electrónicos, recursos de conectividad, servicios web, componentes de análisis de datos, aplicaciones de visualización y servidores de almacenamiento. Una arquitectura general para un sistema de IoT especifica los bloques funcionales que intervienen en el sistema:

- Nodos IoT: dispositivos con sensores / actuadores conectados directamente a través de redes inalámbricas y que acceden a Internet. Cada cosa tiene datos que pueden ser compartidos en internet.

- Gateway: dispositivos que actúan como intermediarios entre las cosas y la nube, proporcionando la conectividad, la seguridad y la capacidad de administración necesarias. Por ejemplo, un punto de acceso Wifi, es una opción popular en implementaciones de monitoreo remoto.

- Plataforma IoT: servicios en línea o infraestructura en la nube, que sirve de colector de los datos provenientes de los nodos y donde se ejecutan servicios para el tratamiento de estos datos.

- Aplicaciones: la forma en que el sistema presenta los datos al usuario final. A veces se puede utilizar un panel de control basado en la web. [7] [8] [9]

## II. SISTEMA IoT PARA MONITOREO DEL ICCA

Este trabajo tiene como objetivo aplicar técnicas de IoT, sistemas integrados y sensores electrónicos en el diseño e implementación de una estación telemétrica de bajo costo para monitorear el contaminante de PM en el entorno circundante. Los datos producidos por esta estación se pueden ver en un panel web accesible a través de Internet. El desarrollo metodológico de esta investigación se basó en el Modelo de referencia Arquitectónico de un sistema IoT. [10].

## A. Proposito y Requerimiestos del sistema

Propósito: el monitoreo automático del ICCA debido a contamínate de material de partículas PM2.5, con comunicación Wifi para envío en tiempo real a través de un sitio web. Comportamiento: dotado de sensores capaces de detectar el nivel de contaminante PM en el entorno circundante conectado a un controlador digital central, programado para realizar una lectura periódica del sensor y transmitir datos a través de Wi-Fi a una plataforma en Internet. Requisito para la gestión: el sistema puede ser monitoreado a través de internet; su gestión y configuración de la programación será local a través de un puerto USB provisto en la propia estación. Requisito para el análisis de datos: el procesamiento de los datos recopilados por el sensor se realiza en la propia estación, la trama de datos se envía al servicio en la "nube". Despliegue de aplicaciones: el control de firmware o software está en la memoria flash del controlador digital dentro de la estación, el monitoreo de los datos producidos por la estación se realiza desde la plataforma IoT configurada para ese propósito. Requisitos de seguridad: el sistema debe tener autenticación de usuario básica para la modificación y acceso a la plataforma IoT, sin embargo, será de acceso público para la visualización de las mediciones.

### B. Especificacion del proceso del sistema

El proceso a ejecutarse en la estación electrónica de monitoreo se define en un bucle repetitivo a través del firmware en el controlador digital: cuando se inicia el sistema, se ejecutan las acciones de configuración para el hardware interno y externo del controlador, luego se activan y leen los sensores para PM2.5, PM10 y Temperatura. Estos datos se formatean y se envían al servicio de IoT a través de la red Wifi local disponible en el sitio de instalación, este proceso se repite periódicamente.

### C. Especificacion del Modelo del sistema

El modelo de dominio de un sistema de IoT define a este como dos entidades de un mismo objeto o dispositivo, una en el medio físico y otra en el medio virtual, para el caso del sistema de monitoreo desarrollado tenemos lo siguiente.

Entidad física: el aire circundante, cuyo nivel de PM contaminante es medido. Entidad virtual: es una representación de la entidad física en el mundo digital, definimos sólo uno para el aire circundante. Dispositivo: controlador digital programable con sensores para contaminante PM y temperatura Recurso: firmware que se ejecuta en el dispositivo y un script que se ejecuta en la nube de IoT. Servicio: servicio ejecutado de forma nativa en el dispositivo.

### D. Especificacion Funcional del Sistema

La vista funcional define grupos funcionales para las diferentes funciones del sistema de IoT. Cada grupo funcional provee, o bien las funciones para interactuar con las instancias de los conceptos definidos en la el modelo de dominio o información relacionada con esos conceptos. En la vista funcional para el sistema IoT de esta propuesta se especifican los siguientes grupos funcionales: Dispositivo: incluye el controlador programable, un sensor de material de partículas PM y un sensor de temperatura. Comunicaciones: incluye los protocolos de comunicación del sistema IoT, para el caso – capa de enlace 802.11, capa de red IPv4, capa de trasporte TCP y capa de aplicación HTTP. La comunicación se realiza por medio de una API basada en RCP. Servicios: Solo se tiene un único servicio ejecutando dentro del sistema IoT, el servicio controlador. Administración: se realiza por medio del recurso de programa o firmware. Seguridad: el mecanismo de seguridad es basa en credenciales de usuario único o "*token*". Aplicación: la interface de monitoreo de los valores producidos por el sistema de IoT se realiza en la "nube" a modo de página web. En la figura 3 se puede observan un esquema que identifica los bloques funcionales del sistema diseñado.

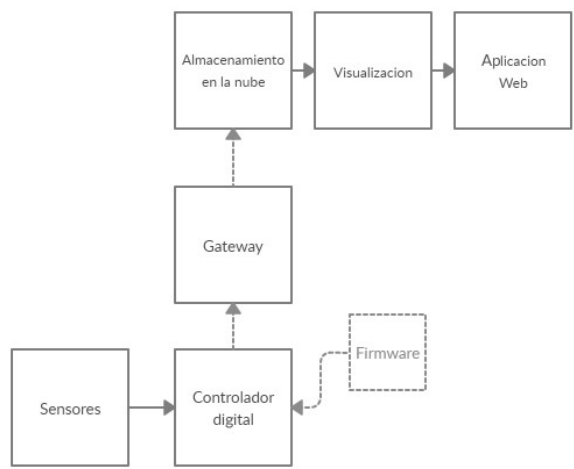

Fig. 3. Especificación de bloques funcionales del sistema de IoT.

### E. Especificacion Operacional del Sistema

Se definen las opciones pertinentes al despliegue y operación del sistema IoT, tales como servicios de alojamiento, almacenamiento, dispositivos, aplicaciones, etc. Nodo IoT: como controlador electrónico programable se usa un microcontrolador Esp32 junto una tarjeta de desarrollo Wemos-Lolin, como elemento sensor de material de partículas se selecciona el sensor PMS5007 y para la magnitud de temperatura se elige el sensor DS18B20, como actuador se dispone de una pantalla Oled para despliegue en el dispositivo de los valores captados. Protocolos de comunicación: 802.11, IPV4/6, TCP y HTTP. Servicios: firmware alojado en el dispositivo, implementado en lenguaje C y ejecutándose como servicio nativo.

### F. Especificacion de la Arquitectura general del Sistema

La plataforma o nube de IoT implementada para el sistema de monitoreo de calidad del aire permite realizar la recepción, análisis y visualización de la información recolectada por las estaciones o Nodos IoT. La arquitectura utilizada, ver figura 4, representa muchas ventajas ya que es escalable para soportar múltiples sensores conectados, dinámica en cuanto a la comunicación entre usuario-plataforma-dispositivos, y segura ya que cuenta con usuarios, *tokens* o claves de acceso, y permisos para la visualización y publicación de paneles de información, dispositivos y análisis de resultados. Para el caso la nube diseñada está compuesta por dos bloques: la plataforma de IoT y la etapa de aplicación denominada Enterprise.

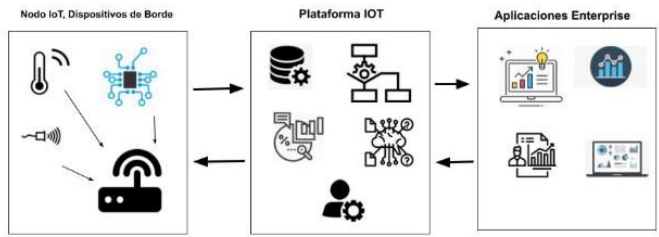

Fig. 4. Arquitectura de IoT implementada

*1) Plataforma IoT*

Es el núcleo de esta arquitectura y es la encargada de diversas partes esenciales en el proceso: ingestión de datos, análisis de datos, administración de dispositivos, alertas y almacenamiento.

Por medio del bloque Plataforma IoT es posible desarrollar flujos de trabajo e información que permitan analizar cada entrada de datos desde los sensores, realizar múltiples operaciones, desde el almacenamiento, copias de información (*backups*), cadena de reglas, análisis de promedios, medianas, máximos y mínimos en los valores recolectados. Lo anterior con el propósito de determinar la emisión de alarmas, de acuerdo a los parámetros previamente establecidos, y/o la ejecución de procesos predeterminados para aplicar acciones correctivas. En el caso de la medición de la calidad del aire, estas alarmas pueden alertar por medio de correo electrónico, SMS, o cualquier otro servicio de mensajería, cuando esta sea Dañina para la salud, previniendo y alertando de esta manera a grupos sensibles, y a las personas en general a utilizar protectores contra estos contaminantes y reducir el riesgo de enfermedades y problemas respiratorios.

*2) Aplicación Eneterprise o Despliegue*

Son las aplicaciones de negocio que utilizan la información generada por la plataforma de IoT para diversos propósitos, desde enriquecer procesos de negocios, reportes, así como el propósito de proporcionar información de apoyo en la toma de decisiones. Las Enterprise apoyada con procesos de Machine Learning o aprendizaje automático, pueden enfocarse además en cambiar flujos de trabajo en la producción a conveniencia de modelos predictivos y proporcionar una retroalimentación a todo el proceso. Lo antes descrito ofrece la oportunidad de generar valor a partir de gráficos funcionales a niveles macros, esto es, generar consolidados que ayuden a tener un parámetro general de la calidad de aire en todo San Salvador y/o por zonas de interés, esto es posible por medio de múltiples sensores en zonas estratégicas.

### G. Integracion de los Componentes del Sistema

En la figura 5 se muestran el diseño del diagrama esquemático o circuito de conexiones eléctricas del dispositivo o nodo de IoT del sistema. Los principales componentes usados son: microcontrolador ESP32 en tarjeta de desarrollo *Wemos Lolin* con pantalla de cristal líquido Oled integrado y los sensores PMS5003 y DS18b20. La conexión entre el sensor de contaminante PM y el microcontrolador se realiza usando el puerto UART. La comunicación con el sensor de temperatura se utilizan los pines del puerto I2C del microcontrolador.

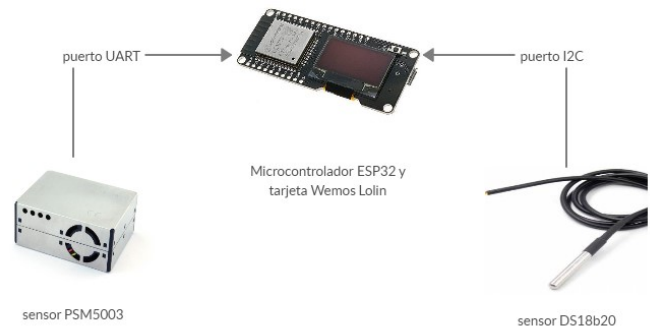

Fig. 5. Integración de principales componentes electrónicos del Nodo IoT.

### H. Desarrollo y Despliegue de Aplicaciones del Sistema

Las aplicaciones o programas que se ejecutan en el sistema de IoT desarrollado son:

- Firmware del microcontrolador: desarrollado en lenguaje Arduino C, en la figura 6 se puede ver su flujograma de control. El programa sigue una estructura cíclica, basada en tareas bien puntuales: 1. Lectura de los sensores para obtener variables con los valores de PM10, PM2.5 y Temperatura, 2. Almacenar estos valores localmente, 3. Mostrar las lecturas en la pantalla local del dispositivo, 4. Formateo de la trama o paquete a enviar a la nube, 5. Envío a la nube, 6. Esperar por la próxima lectura. Cabe mencionar que se enviará las lecturas crudas de las magnitudes y ese en la plataforma IoT que se realiza el cálculo del ICCA. Esto con el objeto de no saturar el microcontrolador con cálculos que se pueden realizar en la nube.

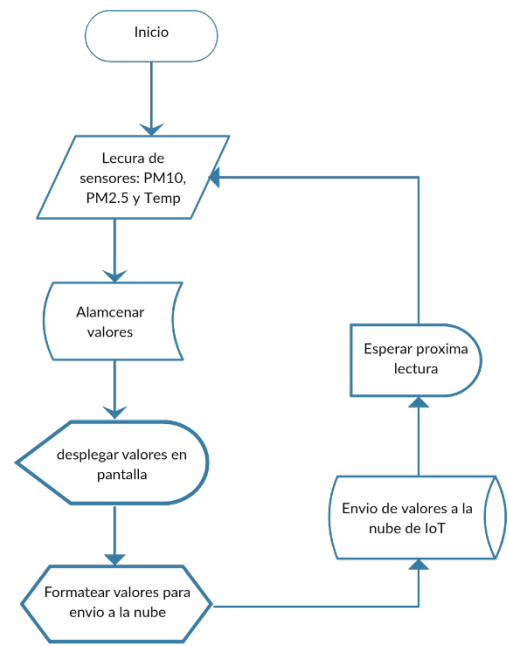

Fig. 6. Flujograma del firmware a implementar en el Nodo IoT.

- Script de servicio en la nube: Desarrollado en lenguaje Java Script, alojado en los servicios de Amazon Web Services (AWS), usando para la conexión la API respectiva. El protocolo telemétrico JavaScript Object Notation (JSON) es utilizado para el envío y recepción de datos entre el Nodo y la Plataforma IoT. Se seleccionó los servicios de AWS por su bajo coste, alta fiabilidad y disponibilidad frente a otros servicios similares. Además de poseer una curva de aprendizaje relativamente corta.

- Aplicación web: al igual que en el punto anterior, se utilizó los servicios de AWS, específicamente hosting para configurar el sitio web de presentación y los tableros o *dashboard* para despliegue de gráficos y tablas con los valores captados por el nodo IoT.

### III. RESULTADOS Y DISCUSION

Luego de la etapa de diseño, se procedió a la construcción e implementación en campo del sistema, obteniendo los resultados que a continuación se presentan.

#### A. Estacion telematica para monitoreo de calidad del aire

El principal resultado de este proyecto es el prototipo de IoT de estación electrónica que permite monitorear los niveles de contaminación por material de partículas, la cual se muestra en la figura 7, y que entre sus características de diseño se pueden mencionar: es un diseño a la medida de las necesidades propias de las condiciones de monitoreo de sistema salvadoreño, se basó en la elección de componentes electrónicos de última generación, así como asequibles y eficientes, su diseño está abierto a la expansión de magnitudes a monitorear. La estación envía a la plataforma IoT los valores de las 3 magnitudes leídas: PM10, PM2.5 y Temperatura cada cierto periodo de tiempo, este valor es configurable vía firmware. La norma nacional sobre lectura del ICCA indica que el cálculo se puede hacer en base a un promedio de las últimas 24 horas.

A continuación, se mencionan detalles de operación del prototipo. Operación eléctrica: Voltaje de operación: 110 Vac. Consumo de potencia: 0.4W Max. Temperatura de trabajo: +60 °C Max. Dimensiones: 15x10x6cm AxLxP. Operación de medición: Rango efectivo PM2.5/10: 0~500 ug/m3. Error máximo: ±10%. Rango efectivo medición temperatura: 0~150 °C. Error máximo: ±5%. Rango de cobertura: 1 litro de aire por toma. Operación de comunicación: Enlace: Internet Wifi 802.11. Potencia de transmisión: 14 dBm típica.

La instalación de las estaciones es sencilla, puede ser empotrada en una pared de un edifico o estructura, con una altura entre 1.5 a 2 metros del nivel del suelo. Los requisitos técnicos que se buscan en el lugar de instalación son: acceso a energía eléctrica Vac y que en la zona exista cobertura de una red Wifi, las estaciones están configuradas para acceso a red por DHCP. La puesta en funcionamiento requiere únicamente definir, vía el firmware, las credenciales de acceso a la red y el tiempo entre reportes a la plataforma IoT, en el caso de estudio se usa 20 minutos.

#### B. Sitio web de Monitoreo y pruebas de campo

El sistema diseñado se implementó en una primera etapa con 5 estaciones o nodos sensores. La primera estación se colocó en el centro de San Salvador, dentro del campus de la Universidad Tecnológica de El Salvador, un sector de la capital de alto tráfico, específicamente entre 19 Avenida y Calle Arce. Este punto fue seleccionado para observar su desempeño y verificar la operación de la operación y para hacer los ajustes necesarios.

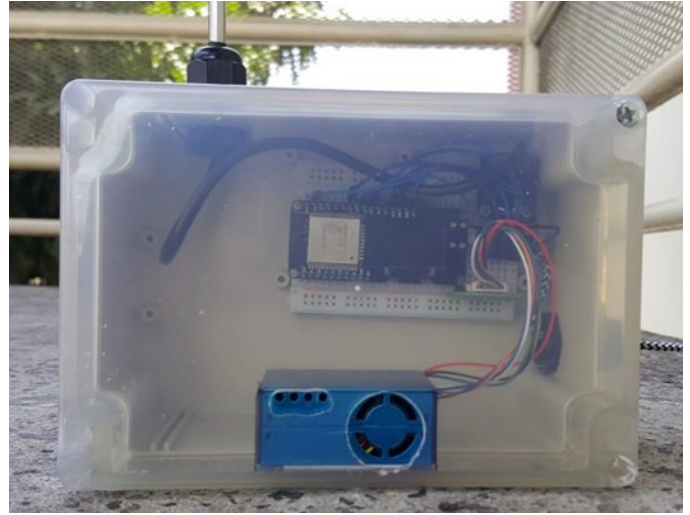

Fig. 7. Prototipo electrónico completo de la estación.

Luego, una segunda estación fue colocada en la ciudad de Santa Ana, aproximadamente a 70 kilómetros al oeste de la capital, esta estación se encuentra en el centro histórico de la ciudad. La tercera estación fue colocada en el centro de la ciudad de Chalatenango al norte de país, igualmente en el centro histórico de la ciudad. La cuarta estación fue colocada en la ciudad de Santa Tecla a 10 km. al oeste de la capital, y esta ubicada en la periferia de la ciudad. La quinta estación se instaló en la ciudad de Antiguo Cuscatlán unos 10 km. al sur de la capital, igualmente en el centro de dicha ciudad. Estos puntos fueron elegidos adrede para poder observar el desempeño en dos lugares con aparente carga de contaminación diferente. Los datos recolectados por las estaciones que hasta el momento conforman el sistema pueden ser consultados, a través del sitio web al que se accede con la dirección siguiente: https://tinyurl.com/utecAQ

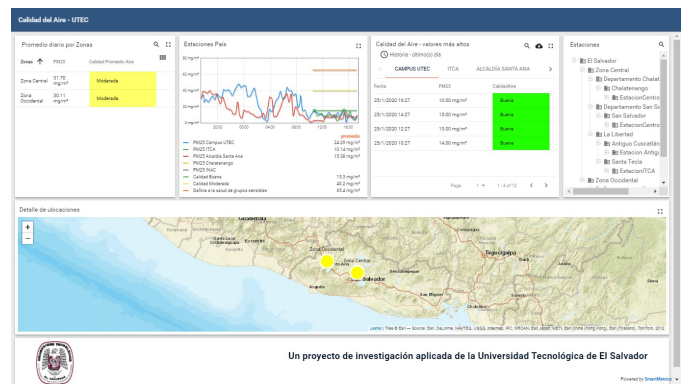

Fig. 8. Pantalla principal del sitio web para monitoreo general de los nodos IoT.

Desde la pantalla principal del sitio de monitoreo, figura 8, se puede observar un resumen general de estado de cada nodo o estación, tanto en los bloques superiores donde observan datos más recientes producidos además de un gráfico sobre puesto con la tendencia de contaminantes reportada por cada nodo. En la parte inferior de sitio principal se puede apreciar, a modo de punto coloreado, el ICCA por estación y su ubicación en el mapa del país, igual al cliquear sobre estos puntos se accede a los datos más puntuales por región del país, y luego al seleccionar el punto de cada estación, se accede a el tablero con información individual de cada una de estas, Figura 9.

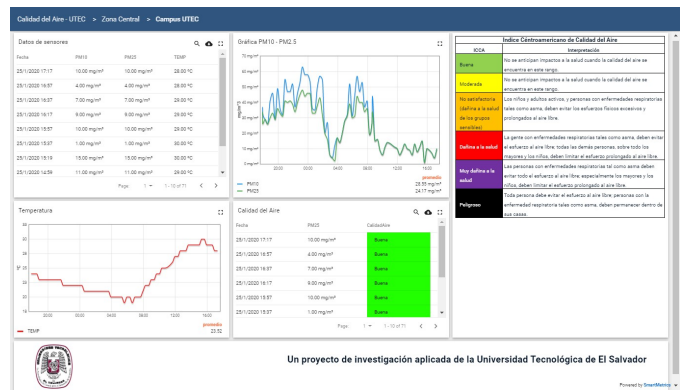

Fig. 9. Pantalla del sitio web con información detallada para cada Nodo IoT.

En esta página web se encuentran información detallada de cada estación, figura 9, como paneles de visualización y tablas de historial de valores de los datos producidos: contaminantes PM10, PM2.5 y Temperatura, además del cálculo e indicación en colores del ICCA en tiempo real la estación.

El sistema completo con las 5 estaciones está operando, se observa que el nivel de contaminación de PM en el sector de San Salvador aumenta durante las horas de alto tráfico vehicular y disminuye durante la noche cuando el tráfico es bajo. Lo cual es un comportamiento similar a las demás estaciones ubicadas en ciudades como Santa Ana y Chalatenango. Las demás estaciones por estas en ciudades de baja concentración poblacional y por ende bajo tráfico vehicular, rara vez sobre pasan el umbral de "moderada" en su calidad de aire.

Se observan efectos significativos en los valores de contaminación en respuesta a las condiciones climáticas como las estaciones de lluvia y viento. Los valores de contaminación disminuyen a sus valores más bajos después de una lluvia o una ráfaga de viento. Los valores de contaminación disminuyen a sus valores más bajos después de una lluvia o una ráfaga de viento.

## IV. CONCLUSIONES Y TRABAJO FUTURO

El desarrollo de una estación de monitoreo junto a la plataforma web para el monitoreo del nivel de contaminante PM en el ambiente, es un paso fundamental en el estudio de comportamiento, impactos y acciones dentro del cuido de medio ambiente.

El prototipo electrónico de la estación de monitoreo, así como la plataforma IoT automatizada, fueron desarrollados utilizando técnicas actuales de electrónica, programación e internet de las cosas, lo cual permitió producir un equipo a bajo costo y que funciona según los requisitos esperados, cumpliendo así el objetivo principal de esta primera fase del proyecto de línea de investigación sobre monitoreo ambiental.

El uso de herramientas como el microcontrolador ESP32 junto con el lenguaje de programación C y el uso de los protocoles de trasferencia basados en JSON permite el desarrollo de prototipos de IoT a un bajo costo, con tiempos de desarrollo cortos y un alto desempeño.

Los datos producidos por las estaciones de monitoreo remoto, han sido comparados con los reportados por estaciones cercanas propiedad del Ministerio de Medio Ambiente y Recursos Naturales, con lo cual se ha verificado que el desempeño es similar.

Se ha aportado conocimiento científico nuevo, de manera que se muestran nuevas e innovadoras técnicas de uso de componentes de hardware y software, en la implementación de sistemas de internet de las cosas. Estas pueden ser aplicadas en nuevos desarrollos, lo que permitiría la creación de prototipos de forma rápida y eficiente.

A futuro, la presente investigación tiene como trabajo el desarrollo de más estaciones para diferentes locaciones dentro del territorio nacional, realizar experimentos de validación en conjunto con el Ministerio de Medio Ambiente y Recursos Naturales, implementar una red de monitoreo a través de enlaces de radiofrecuencia, el análisis de datos masivos producidos por las estaciones futuras. Utilizar los conocimientos frutos de este proyecto en el desarrollo de nuevas líneas de investigación aplicada, en áreas como: análisis de mantos acuíferos, monitoreo en campos de agricultura y ganadería, análisis de desempeño deportivo, etc.